# Physical Vapor Transport Growth of Antiferromagnetic CrCl$_3$ Flakes Down to Monolayer Thickness

Jia Wang[1], Zahra Ahmadi[2], David Lujan[3], Jeongheon Choe[3], Takashi Taniguchi[4], Kenji Watanabe[5], Xiaoqin Li[3], Jeffrey E. Shield[2], and Xia Hong[1*]

[1] Department of Physics and Astronomy & Nebraska Center for Materials and Nanoscience, University of Nebraska-Lincoln, Lincoln, NE 68588-0299, USA

[2] Department of Mechanical and Materials Engineering, University of Nebraska-Lincoln, Lincoln, NE 68588-2526, USA

[3] Department of Physics, University of Texas at Austin, Austin, TX 78712-1192, USA

[4] International Center for Materials Nanoarchitectonics, National Institute for Materials Science, 1-1 Namiki, Tsukuba 305-0044, Japan

[5] Research Center for Functional Materials, National Institute for Materials Science, 1-1 Namiki, Tsukuba 305-0044, Japan

[*] Corresponding E-mail: xia.hong@unl.edu

The van der Waals magnets CrX$_3$ (X = I, Br, and Cl) exhibit highly tunable magnetic properties and are promising candidates for developing novel two-dimensional (2D) spintronic devices such as magnetic tunnel junctions and spin tunneling transistors. Previous studies of the antiferromagnetic CrCl$_3$ have mainly focused on mechanically exfoliated samples. Controlled synthesis of high quality atomically thin flakes is critical for their technological implementation but has not been achieved to date. This work reports the growth of large CrCl$_3$ flakes down to monolayer thickness via the physical vapor transport technique. Both isolated flakes with well-defined facets and long stripe samples with the trilayer portion exceeding 60 μm have been obtained. High-resolution transmission electron microscopy studies show that the CrCl$_3$ flakes are single crystalline in the monoclinic structure, consistent with the Raman results. The room temperature stability of the CrCl$_3$ flakes decreases with decreasing thickness. The tunneling magnetoresistance of graphite/CrCl$_3$/graphite tunnel junctions confirms that few-layer CrCl$_3$ possesses in-plane magnetic anisotropy and Néel temperature of 17 K. This study paves the path for developing CrCl$_3$-based scalable 2D spintronic applications.

Since their discovery, two-dimensional (2D) van der Waals (vdW) magnets CrX$_3$ (X=Cl, Br, I) have attracted extensive research interests for their unusual magnetic properties[1-10] compared with conventional magnetic metals and oxides.[11-13] They are flexible, can sustain the magnetic ground state down to monolayer thickness,[6-7, 9-10] and can be stacked with other vdW materials to create multifunctional heterostructures.[1-6, 14-18] It has been shown that the magnetic order and magnetic anisotropy of CrX$_3$ can be sensitively tuned by strain and doping,[16-19] making it a versatile playground for studying magnetic quantum phase transitions and designing novel energy-efficient spintronic devices, including magnetic tunnel junctions,[1-5] spin tunneling field-effect transistors,[16-18] and quantum spin Hall systems.[15] CrCl$_3$ is an A-type antiferromagnet with in-plane magnetic anisotropy.[1-4] Previous studies have mainly focused on mechanically exfoliated samples.[1-5, 8, 20-21] While nanosheets of CrCl$_3$ have been deposited via the chemical vapor transport (CVT) method, only samples thicker than 25 nm have been obtained.[22] Controlled synthesis of high-quality atomically thin flakes is of great fundamental and technological interests but has not been achieved to date.

In this work, we report the direct growth of large CrCl$_3$ flakes down to monolayer thickness via the physical vapor transport (PVT) technique. Triangular and hexagonal thin flakes with well-defined facets as well as long stripes of tri-layer samples with the trilayer portion exceeding 60 μm have been obtained. High-resolution transmission electron microscopy (HRTEM) studies show that the CrCl$_3$ flakes are single crystalline with the monoclinic structure, consistent with the Raman characterizations. The sample stoichiometry has been confirmed by scanning electron microscopy (SEM)-energy dispersive X-ray spectroscopy (EDS) studies. Atomic force microscopy (AFM) studies show that the room temperature stability of CrCl$_3$ flakes decreases with decreasing thickness. Characterization of graphite/CrCl$_3$/graphite tunneling devices reveals a Néel temperature ($T_N$) of 17 K and in-plane magnetic anisotropy in few-layer CrCl$_3$. Our study enables scalable synthesis of high-quality atomically thin CrCl$_3$ flakes, paving the path for their implementation in 2D spintronic applications.

2D vdW CrCl$_3$ flakes are synthesized from CrCl$_3$ powder using the PVT technique (**Figure 1**a, see Experimental Section for growth details). The samples are deposited on three types of substrates: mica (fluorophlogopite, [KMg$_3$(AlSi$_3$O$_{10}$)F$_2$]), highly oriented pyrolytic graphite (HOPG), and SiO$_2$/Si substrates. We then investigate the effects of substrates on the lateral size, flake thickness, and crystalline orientation of the samples. As shown in Figure 1b, CrCl$_3$ on SiO$_2$/Si prefers vertical growth and forms relatively thick crystals. Horizontal growth of large size thin flakes has been achieved on HOPG (Figure 1c) and mica (Figure 1d) substrates, which can be attributed to their atomically smooth and dangling bond-free surfaces. Previous studies have shown that such surfaces can effectively promote the attachment of precursors on the layer edges and facilitate subsequent horizontal growth.[23-24] As the flakes deposited on HOPG do not have well-defined facets (Figure 1c) and are hard to isolate from the underneath graphite pieces, we next focus on characterizing the samples deposited on mica.

We have obtained both isolated flakes with triangular and hexagonal shapes and long stripe samples on mica. Figures 1e-i show the AFM topography images of five CrCl$_3$ samples with different thicknesses. The flakes thicker than 9 nm show well-defined facets with sharp edges (Figures 1e-f). The few-layer to monolayer CrCl$_3$ flakes (Figures 1g-h) also possess the triangular shape, but their edges are rough with micro-facets and the corners are rounded. This has been attributed to the CrCl$_3$ desorption during growth. For ultrathin flakes, there is an insufficient growth time for the edge atoms to reach thermodynamic equilibrium.[25] In previous studies, CVT-grown CrCl$_3$ nanosheets are mostly thicker than 25 nm,[22] and ultrathin flakes have only been obtained via mechanical exfoliation.[1-5, 20] Our study is the first report of direct growth of monolayer CrCl$_3$ (Figure 1h). Systematic AFM imaging over a large scale reveals about 25% yield of ultrathin flakes, including monolayer, bilayers, and trilayer samples (Figure S1a-b, Supporting Information). In addition to the isolated flakes, we have also achieved long stripes of ultrathin CrCl$_3$ samples. Figure 1i shows the trilayer portion (66 μm by 20 μm) of a long stripe sample, whose overall length is over 1 mm (Figure S1c, Supporting Information). The ultrathin portion of the stripe samples can exceed 60% (Supporting Information Section 1).

We examine the room temperature stability of the CrCl$_3$ flakes by taking a series of AFM images after growth.[20, 26] It has been shown that CrCl$_3$ are more stable compared with CrI$_3$.[20-21, 26] For a 64 nm flake, there is no obvious change in the sample morphology for about 5 months (Figure S2a, Supporting Information), showing excellent room temperature stability. The thinner flakes, on the other hand, show clear degradation with time. The 20 nm flake remains stable on Day 23, while bubble-like features emerge on the sample surface on Day 37 (Figure S2b, Supporting Information). Similar degradation signs have been reported in exfoliated CrCl$_3$ flakes,[20, 26] which is attributed to the formation of CrCl$_2$.[26] For the monolayer flakes, the sample surface becomes rough on Day 6 (Figure S2c, Supporting Information), with the flake thickness increasing from 0.69 nm to 3.05 nm (Figure S2d, Supporting Information). It is possible that the sample degradation has started even before it becomes discernable in AFM measurements, as previously reported in Cr$_2$Ge$_2$Te$_6$.[27]

We carry out TEM, SEM, and Raman measurements to characterize the sample structure and stoichiometry. Our studies show that $CrCl_3$ can be easily damaged when exposed to electron beam and laser excitation (Figure S3, Supporting Information). To ensure the data quality, we have reduced the exposure time and used minimal laser power in these measurements and focused on characterizing relatively thick samples (> 20 nm). At room temperature, $CrCl_3$ possesses the monoclinic structure, which belongs to the C2/m space group (**Figure 2**a).[28] The Cr atoms form a honeycomb structure in the *a-b* plane, with each Cr atom surrounded by the Cl octahedron. Figure 2b shows an HRTEM image of $CrCl_3$, where the crystalline planes of (020), ($1\bar{1}0$), ($\bar{1}\bar{1}0$) make a quasi-equilateral triangle. The inter-planar spacing *d* is about 5.1 Å, agreeing with the expected lattice parameter for the monoclinic structure. The corresponding selected-area diffraction (SAD) pattern (Figure 2c) is also consistent with the monoclinic crystal structure.[28] The sharp diffraction peaks and the absence of impurity phases confirm the high crystallinity of the sample.

The stoichiometry of the sample is investigated using SEM-EDS (Figure 2d-f). Element mapping of the Cr K-line (Figure 2e) and Cl K-line (Figure 2f) reveals a homogeneous distribution. From the EDS spectrum, we extract a Cr/Cl ratio of 0.304 (Figure S4, Supporting Information), reasonably close to the ideal ratio of 1/3 considering the uncertainties related to EDS quantitative analysis, as there is significant background signal from the underlying substrate for thin film samples.[22] No signal of sulfur is detected in the EDS spectrum, confirming that the sample purity is not affected by the S powder used for promoting sample nucleation.

Next, we carry out Raman studies at room temperature. To minimize the sample damage by laser heating,[29] we transfer the samples onto Au-coated $SiO_2$/Si substrates to facilitate energy dissipation. **Figure 3**a shows the Raman spectra of $CrCl_3$ flakes with different thicknesses. For micron-thick bulk samples, we observed six Raman peaks at about 123 cm$^{-1}$, 165 cm$^{-1}$, 207 cm$^{-1}$, 244 cm$^{-1}$, 300 cm$^{-1}$, and 344 cm$^{-1}$, which are denoted as $A_g(1)$, $A_g(2)$, $A_g(3)/B_g$, $A_g(4)$, $A_g(5)$, and $A_g(6)$ modes, respectively. The spectrum agrees with the monoclinic structure of $CrCl_3$.[3, 20] The Raman signal decreases with sample thickness and becomes hard to resolve in flakes thinner than 20 nm. For the signal that can be detected, there is no noticeable peak shift with flake thickness.

Figure 3b shows the polarized Raman spectra taken on a 43 nm $CrCl_3$ flake. Compared with the polarized Raman spectra of bulk $CrCl_3$ crystal,[30] we only resolve five $A_g$ phonon modes in the parallel polarization (XX) and one $B_g$ mode in the perpendicular polarization (XY) due to the relatively low signal strength in thin flakes. The peak position for the $B_g$ mode (about 207 cm$^{-1}$) contains two modes $B_g(3/4)$ with degenerate energy.[30] Figure 3c shows the polar maps of XX Raman intensity for the four $A_g$ modes with relatively high intensity, where the angle of the incident light polarization $\theta$ is defined with respect to the $a$-axis of $CrCl_3$ (Figure S5, Supporting Information). All $A_g$ modes exhibit two-fold symmetry, with four local maxima occurring at $\theta = 0°, 90°, 180°, 270°$. The intensity at 0° and 180° is higher than that at 90° and 270°. The Raman intensity is proportional to $|\boldsymbol{g}_s \tilde{R} \boldsymbol{g}_i^T|^2$, where $\boldsymbol{g}_i$ ($\boldsymbol{g}_s$) is the polarization vector of the incident (scattered) light and $\tilde{R}$ is the Raman tensor.[31] In the XX configuration, $\boldsymbol{g}_s = \boldsymbol{g}_i \propto (\cos\theta, \sin\theta, 0)$. For a monoclinic structure, the angular-dependent Raman response in XX is given by: $I(A_g) \propto |a\cos^2\theta + b\sin^2\theta|^2$ and $I(B_g) \propto e^2 \sin^2(2\theta)$, where $a$, $b$, and $e$ are fitting parameters.[31] Previous studies have shown that both $A_g$ and $B_g$ modes can contribute to the polar mapping,[31] so the overall Raman intensity can be expressed as:

$$I' \propto |a\cos^2\theta + b\sin^2\theta|^2 + e^2 \sin^2(2\theta). \tag{1}$$

As shown in Figure 3c, Equation (1) well describes the angular dependence of Raman intensity, further confirming that $CrCl_3$ is crystallized in the monoclinic structure.

To probe the magnetic properties of the sample, we fabricate few-layer $CrCl_3$ into tunnel junction devices (**Figure 4**a) and characterize their tunneling magnetoresistance (TMR). Figure 4b shows a device composed of a 6-layer CrCl$_3$ tunnel barrier (Figures S6-S7, Supporting Information) sandwiched between top and bottom thin graphite flakes transferred on a SiO$_2$ substrate with prepatterned gold electrodes (Experimental Section). The effective area of the tunnel junction is about 10.8 μm$^2$. The device is then encapsulated by a top h-NB flake to avoid ambient degradation. At room temperature, the *I-V* characteristic of the device is highly stable for over 2 months in the ambient conditions, which is the duration of measurements (Figure S8, Supporting Information).

Figure 4c shows the tunneling characteristic of the device at various temperatures. The tunneling current decreases rapidly with decreasing temperature below 300 K and exhibits weak temperature-dependence below 50 K. Plotting *I*/*V*$^2$ versus 1/*V* reveals two distinct regimes, which can be understood by considering the evolution of the dominating tunneling mechanism. At low bias *V* << Φ/*e*, where Φ is the tunnel barrier height and *e* is the elementary charge, the tunneling current is dominated by the direct tunneling mechanism, with the tunneling current given by:[32-33]

$$I \propto V e^{\left(-\frac{2d\sqrt{2m^*\Phi}}{\hbar}\right)}. \tag{2}$$

Here $m^*$ is the effective mass, $\hbar$ is the reduced Plank constant, and *d* is the thickness of the CrCl$_3$ flake. At *V* > Φ/*e*, the Fowler-Nordheim (FN) mechanism becomes dominant, and the current can be expressed as: [32-33]

$$I \propto V^2 e^{\left(-\frac{4d\sqrt{2m^*\Phi^3}}{3\hbar eV}\right)}. \tag{3}$$

Equations (2)-(3) can well capture the data shown in Figure 4c. The transition voltage between these two regimes decreases with increasing temperature, illustrating the enhanced contribution of thermo-carriers tunneling through the bias-modified tunnel barrier.[34]

We then use the transition between the direct and FN tunneling regimes at low temperature to estimate the tunnel barrier height Φ.[32, 35] In Figure 4d, we plot $\ln\left(\frac{I}{V^2}\right)$ *vs.* 1/*V* at 2 K and superimposed the fitting curves for the direct tunneling regime, *i.e.*, $\ln\left(\frac{I}{V^2}\right) \propto \ln\left(\frac{1}{V}\right)$ (Equation (2)), and the FN regime, *i.e.*, $\ln\left(\frac{I}{V^2}\right) \propto \frac{1}{V}$ (Equation (3)). The transition voltage $V_t$ is defined as the crossing point of these two behaviors ($V_t \approx 0.51$ V), which has been used to estimate the height of the tunnel barrier. As the transition is relatively broad, this can lead to about 10% uncertainty in the estimated $\Phi$. Assuming $\Phi = eV_t = 0.51$ eV and considering the layer number of the flake to be 6±1 (Figure S7, Supporting Information), we extract the effective mass for the CrCl$_3$ tunnel barrier to be $m^* = (0.5 \pm 0.1)m_0$, where $m_0$ is the free electron mass.[36]

Figure 4e shows the tunneling *I-V* relation at 2 K with and without a perpendicular magnetic field $B_\perp$. Applying a magnetic field increases the tunneling current, which can be attributed to spin alignment in CrCl$_3$ induced by the magnetic field. Without the magnetic field, the spins in the adjacent layers are antiparallel to each other, which suppresses the electron tunneling probability, yielding an effectively higher tunnel barrier height. An applied field of 6 T can align the spins of all layers along the out-of-plane direction, resulting in higher *I*. At $V = 0.8$ V, the tunneling current changes from 8.4 nA at 0 T to 25.6 nA at 6 T, corresponding to a TMR (6 T) = $100\% \times \frac{I(6\text{ T}) - I(0\text{ T})}{I(0\text{ T})}$ = 206%, which is significantly higher than that obtained on bilayer and trilayer CrCl$_3$ tunneling devices at this temperature in previous experiments.[5] The enhanced TMR shows that the spin filtering efficient increases with increasing barrier thickness.[2]

From the temperature-dependence of zero field tunneling current and its derivative *dI/dT* (**Figure 5**a), we identify a clear kink at 17 K, which corresponds to the $T_N$. The $T_N$ value is consistent with previous reports of bulk[21] and exfoliated CrCl$_3$.[1-2, 4-5] Below and above $T_N$, the tunneling current exhibits distinct magnetic field dependence. As shown in Figure 5b, at 2 K, *I* rises rapidly with increasing magnetic field and saturates around $B_\perp = 2.5$ T. Below $T_N$, the magnetic field aligns the in-plane, anti-aligned interlayer spins to the out-of-plane orientation, which yields higher tunneling current.[1-2, 4-5] Once the spins are fully aligned, increasing the magnetic field no longer changes the tunneling current. At 17 K, in contrast, the tunneling current exhibits a weaker magnetic field dependence and does not saturate up to 6 T. Above $T_N$, the spins do not have long-range order and are randomly oriented. The magnetic field is thus not sufficient to fully align the spins. This change is also reflected in the temperature-dependence of TMR at 6 T (Figure 5b inset), which decreases monotonically with increasing temperature and exhibits a deflection point around $T_N$. We also note that the change of tunneling current below $T_N$ is gradual, in contrast to the sharp change observed in CrI$_3$.[6] This is consistent with the weak in-plane magnetic anisotropy for CrCl$_3$, where the out-of-plane magnetic field induces continuous spin rotation rather than directly flipping the spin orientation.[1-2, 4-5, 21]

In conclusion, we have successfully synthesized large CrCl$_3$ flakes down to monolayer thickness using the physical vapor transport technique, with high crystallinity and homogeneous chemical composition achieved. With h-BN encapsulation, few-layer CrCl$_3$-based tunneling devices exhibit high ambient stability for more than 2 months. The tunneling magnetoresistance reveals that few-layer CrCl$_3$ flakes possess a Néel temperature of 17 K, in-plane magnetic anisotropy, and tunneling magnetoresistance of >200% below $T_N$. Our study enables the direct growth of large size atomically thin CrCl$_3$ flakes, paving the path for implementing this material for scalable 2D spintronic applications.

**Experimental Section**

**Synthesis.** High-quality 2D vdW $CrCl_3$ flakes were deposited in a horizontal single-zone furnace (Thermo Scientific TF55035-A1) with a 1 inch diameter quartz tube by the PVT technique. A quartz boat with $CrCl_3$ source powder (99.99%, Alfa Aesar) was placed at the center of the single-zone tube furnace. A small amount of sulfur powder (99.9995%, Alfa Aesar) was loaded in the upstream of the tube to facilitate sample nucleation. The substrate was placed in the tube at about 10 cm downstream from the $CrCl_3$ source powder. Three types of substrates, mica (highest grade V1 mica disc, MIT), HOPG (Grade 3, SPI), and $SiO_2$/Si were investigated. Before growth, the system was purged by Ar gas three times. During sample growth, the furnace was heated up to 700-750 ºC at a rate of 30 °C min$^{-1}$ with 40 standard cubic centimeters per minute (sccm) Ar process gas, and the tube was kept at one atmosphere pressure. After 5 min growth, the furnace was cooled down to room temperature naturally.

**Sample Characterizations.** The thickness and surface morphology of as-grown $CrCl_3$ flakes were characterized by AFM (Bruker Multimode 8) with the tapping mode. SEM was performed using an FEI Helios Nanolab 660 with a field emission gun at 2 kV. The chemical element analysis was conducted by EDS using the point and mapping modes in SEM. HRTEM studies were performed in an FEI Tecnai Osiris electron microscope operated at 200 kV. Nonpolarized Raman spectra were collected by a Thermo Scientific DXR Raman microscope with a 532 nm laser, a 100x objective, exposure time of 30 s, 0.2 mW laser power, and a 900 lines/mm grating. Polarized Raman spectra were recorded using a Harina/Princeton Acton 7500i/spectrometers equipped with a 532 nm laser, with a 50x objective, 0.2 mW incident laser power, integration time of 20 min, and 1800 lines per mm grating. The excitation laser and collected Raman signal were collinearly polarized. For the angular dependence measurements, the angle step was 5 degrees for a half-wave plate, which was 10 degrees in the polar map. For SEM, TEM, and Raman characterizations, the $CrCl_3$ flakes were transferred onto Au-deposited (10 nm) SiO$_2$/Si substrate (SEM and Raman) and TEM chips (Silicon Nitride Support Film, 50 nm with 0.5x0.5mm Window) using the all-dry stamping transfer technique.

**Device Fabrication and Electrical Characterizations.** Au/Cr (20/5 nm) electrodes were prepatterned into two-point geometry on the SiO$_2$/Si substrates using photolithography followed by e-beam evaporation. The tunnel junction devices were assembled by the all-dry stamping transfer method, which was performed on an optical microscope equipped with a stamping stage. The as-grown CrCl$_3$ flakes were picked up from the mica substrate by an elastomeric film (Gel-Film® WF × 4 1.5 mil from GelPak), which was adhered with a glass slide fixed on the stamping stage. The thin graphite electrodes and the h-BN protection layer were mechanically exfoliated. The graphite, few-layer CrCl$_3$, and h-BN flakes were picked up sequentially by gel-films and stacked into h-BN encapsulated graphite/CrCl$_3$/graphite heterostructures on top of the prepatterned SiO$_2$/Si substrates (Section 6, Supporting Information). The electrical measurements were carried out in a Quantum Design PPMS using an external Keysight 1500A Semiconductor Device Parameter Analyzer.

**Acknowledgements**

The authors would like to thank Qiuchen Wu, Bingqiang Wei, Alexey Lipatov, and Alexander Sinitskii for their technical support. This work was supported by NSF through Grant Nos. DMR-2118828, DMR-1710461, and OIA-2044049. T.T. acknowledges support from the JSPS KAKENHI (Grant Nos. 19H05790 and 20H00354) for h-BN growth. The research was performed, in part, in the Nebraska Nanoscale Facility: National Nanotechnology Coordinated Infrastructure, the Nebraska Center for Materials and Nanoscience, and the Nanoengineering Research Core Facility, which are supported by NSF ECCS: 2025298, and the Nebraska Research Initiative.

**Figures**

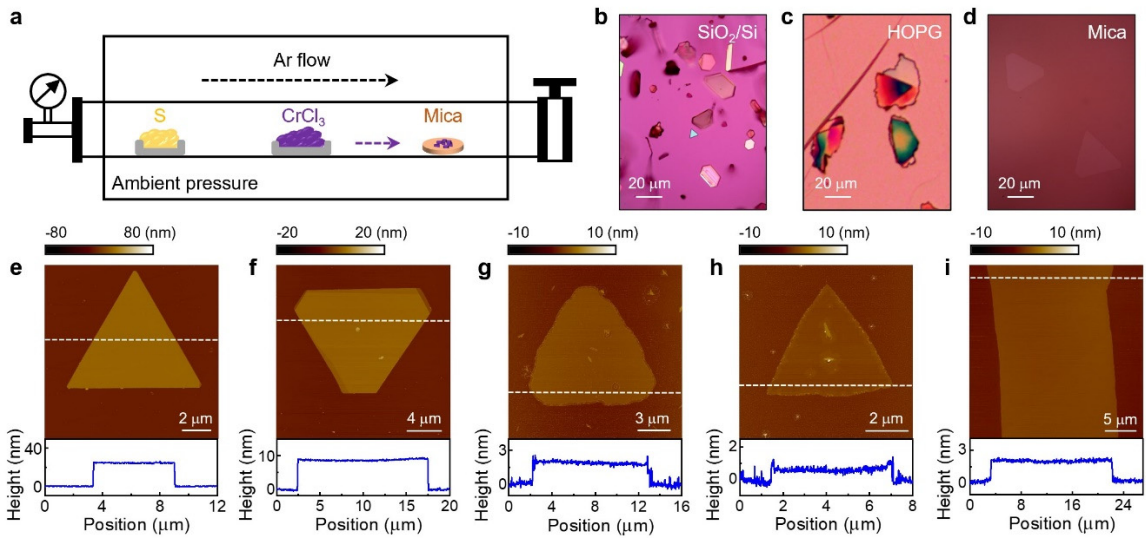

**Figure 1.** Synthesis of thick to monolayer $CrCl_3$ flakes. a) Schematic of the experimental setup for PVT growth of $CrCl_3$. b-d) Optical images of as-grown $CrCl_3$ flakes on (b) $SiO_2/Si$, (c) HOPG, and (d) mica substrates. e-i) AFM images of $CrCl_3$ flakes on mica with different thicknesses (upper panels), with the height profiles along the dashed lines (lower panels). The averaged flake thicknesses are 24.8±0.2 nm, 9.25±0.04 nm, 1.84±0.01 nm (tri-layer), 0.63±0.07 nm (monolayer), and 1.9±0.1 nm (tri-layer), respectively.

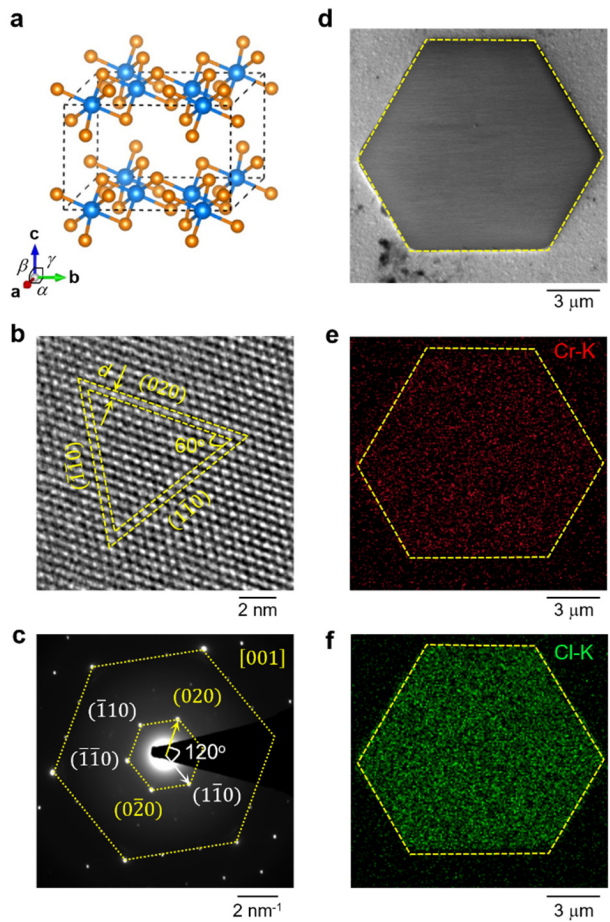

**Figure 2.** Structural characterization and element analysis of CrCl$_3$ flakes. a) Schematic unit cell of monoclinic CrCl$_3$, with $a$ = 5.959 Å, $b$ = 10.321 Å, $c$ = 6.114 Å, $\alpha = \gamma = 90°$, and $\beta = 108.49°$. b) HRTEM micrograph and c) SAD pattern taken on a thick CrCl$_3$ flake along [001] zone axis. d) SEM image of a thick CrCl$_3$ flake, with element mapping of e) Cr and f) Cl.

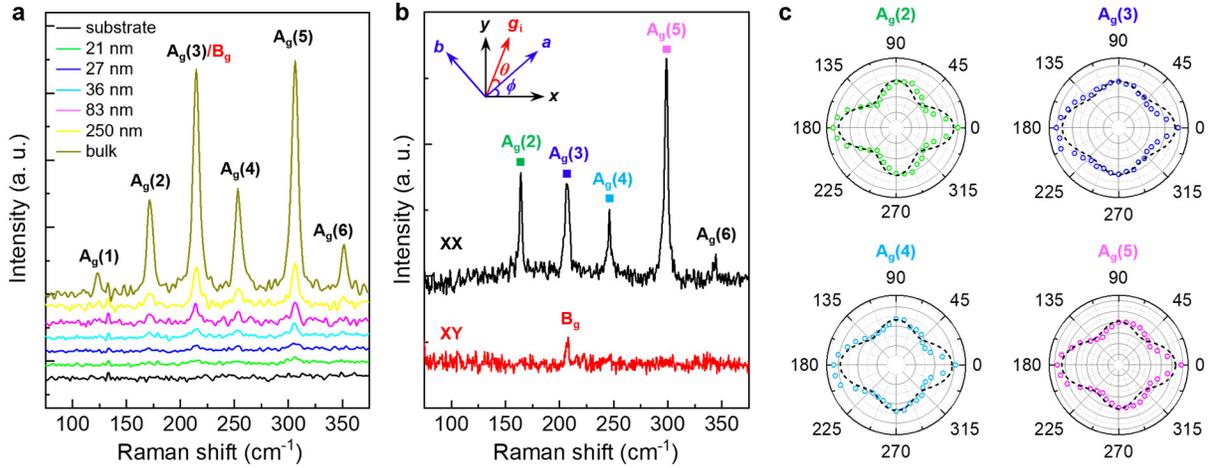

**Figure 3.** Raman characterizaiton of CrCl$_3$ flakes. a) Raman spectra of CrCl$_3$ flakes with different thicknesses. b) Raman spectra of a 43 nm CrCl$_3$ flake measured in parallel (XX) and perpendicular (XY) configurations. Inset: Schematic of crystalline orientations and laboratory coordinates, where the relative angle $\phi$ between $x$-axis and sample $a$-axis is arbitrary. c) Polar plots of integrated Raman intensity for different A$_g$ modes of the same CrCl$_3$ flake shown in (b).

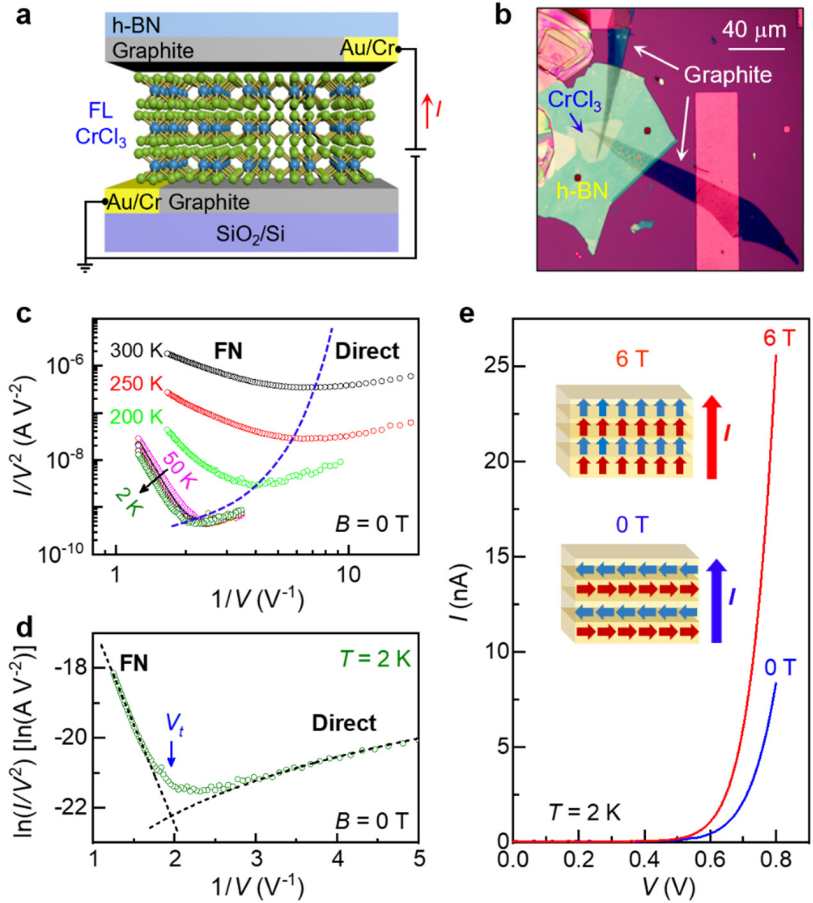

**Figure 4.** Tunneling characteristic of a graphite/6-layer CrCl$_3$/graphite device encapsulated with h-BN. a) Device schematic. b) Optical image. c) Zero field $I/V^2$ vs. $1/V$ at 300, 250, 200, 50, 25, 22, 20, 18, 12 and 2 K. The dashed line serves as a guide to the eye. d) Zero field $\ln(I/V^2)$ vs. $1/V$ at 2 K with fits to Equations (2)-(3) (dashed lines). e) Tunneling $I$-$V$ at 2 K with $B_\perp = 0$ and 6 T. Inset: Schematic of spin orientation in CrCl$_3$ with and without magnetic field.

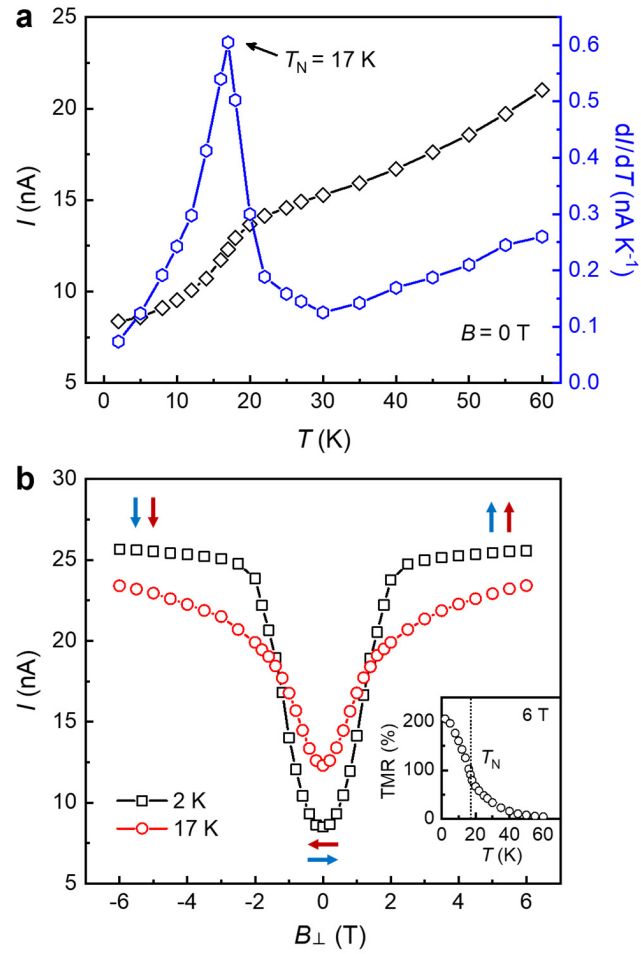

**Figure 5.** Tunneling magnetotransport of the 6-layer CrCl$_3$ tunnel junction device. a) Temperature-dependent tunneling current at zero magnetic field. b) Tunneling current *vs.* $B_\perp$ at 2 K and 17 K. $V$ = 0.8 V. Inset: TMR ratio versus $T$ at $B_\perp$ = 6 T. The dotted line marks $T_N$.

# Supporting Information

**Physical Vapor Transport Growth of Antiferromagnetic CrCl$_3$ Flakes Down to Monolayer Thickness**

Jia Wang[1], Zahra Ahmadi[2], David Lujan[3], Jeongheon Choe[3], Takashi Taniguchi[4], Kenji Watanabe[5], Xiaoqin Li[3], Jeffrey E. Shield[2], and Xia Hong[1*]

[1] Department of Physics and Astronomy & Nebraska Center for Materials and Nanoscience, University of Nebraska-Lincoln, Lincoln, NE 68588-0299, USA

[2] Department of Mechanical and Materials Engineering, University of Nebraska-Lincoln, Lincoln, NE 68588-2526, USA

[3] Department of Physics, University of Texas at Austin, Austin, TX 78712-1192, USA

[4] International Center for Materials Nanoarchitectonics, National Institute for Materials Science, 1-1 Namiki, Tsukuba 305-0044, Japan

[5] Research Center for Functional Materials, National Institute for Materials Science, 1-1 Namiki, Tsukuba 305-0044, Japan

[*] Corresponding E-mail: xia.hong@unl.edu

**Contents:**
1. Thickness Distribution of CrCl$_3$ Samples
2. Room Temperature Stability of CrCl$_3$ Flakes
3. Sample Damage upon TEM and Raman Measurements
4. Element Analysis
5. Polarized Raman Analysis of Crystalline Orientation
6. Assembly of h-BN Encapsulated Graphite/CrCl$_3$/Graphite Tunnel Junction
7. Thickness of CrCl$_3$ Tunnel Barrier
8. Ambient Stability of h-BN Encapsulated Few-Layer CrCl$_3$ Tunnel Junction

## 1. Thickness Distribution of CrCl$_3$ Samples

**Figure S1**a shows an optical image of isolated CrCl$_3$ flakes on a mica substrate. It is challenging to identify the ultrathin flakes due to the poor optical contrast for CrCl$_3$ on mica. We have performed systematic atomic force microscopy (AFM) imaging to search for ultrathin flakes over this area and identified 8 ultrathin samples, including monolayer (1L), bilayer (2L), and trilayer (3L) flakes, out of a total of 30 flakes (Figure S1a-b), corresponding to about 25% yield. Figure S1c shows an optical image of a long stripe sample with over 1 mm length. Based on the AFM measurements (Figure S1d), we estimate that about 60% of the stripe sample is 3L thick.

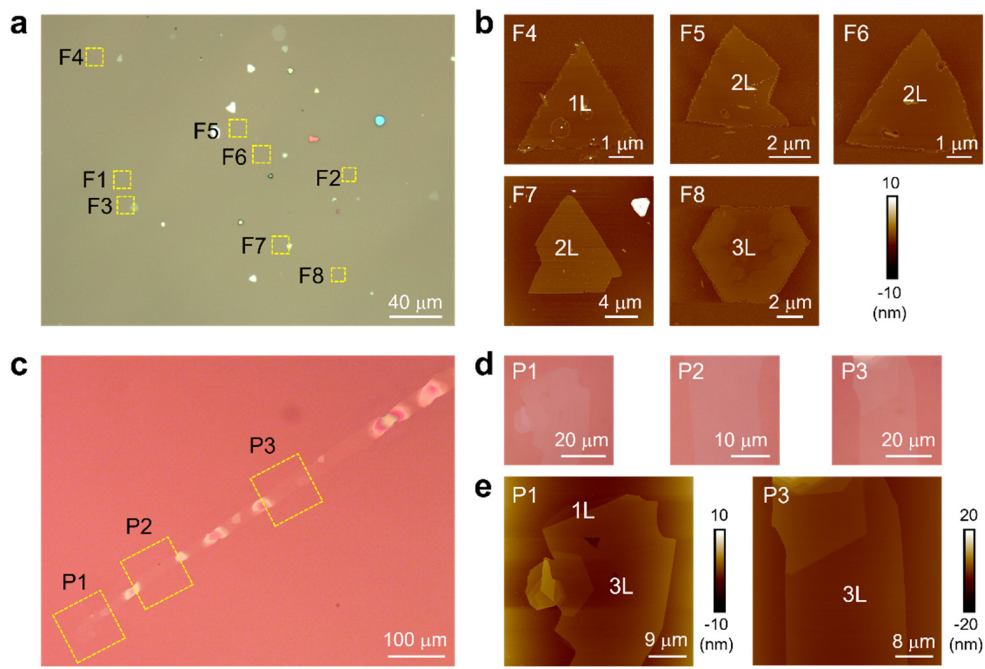

**Figure S1**. Thickness distribution of CrCl$_3$ samples. a) Optical image of as-grown CrCl$_3$ flakes on mica. The AFM images of flakes F1-F3 are shown in Figure 1h, Figure 1g, and Figure S2c, respectively. b) AFM images of CrCl$_3$ flakes F4-F8 marked in (a) (dashed boxes). c) Optical image of a long stripe CrCl$_3$ sample on mica. d) Expanded optical images of areas P1-P3 marked in (c) (dashed boxes). e) AFM images taken on the areas P1 and P3. The AFM image of P2 is shown in Figure 1i in the main text.

## 2. Room Temperature Stability of CrCl$_3$ Flakes

**Figure S2** shows the AFM images of CrCl$_3$ flakes with various thicknesses taken at different time. The stability of the flakes is discussed in Section 2.1 in the main text.

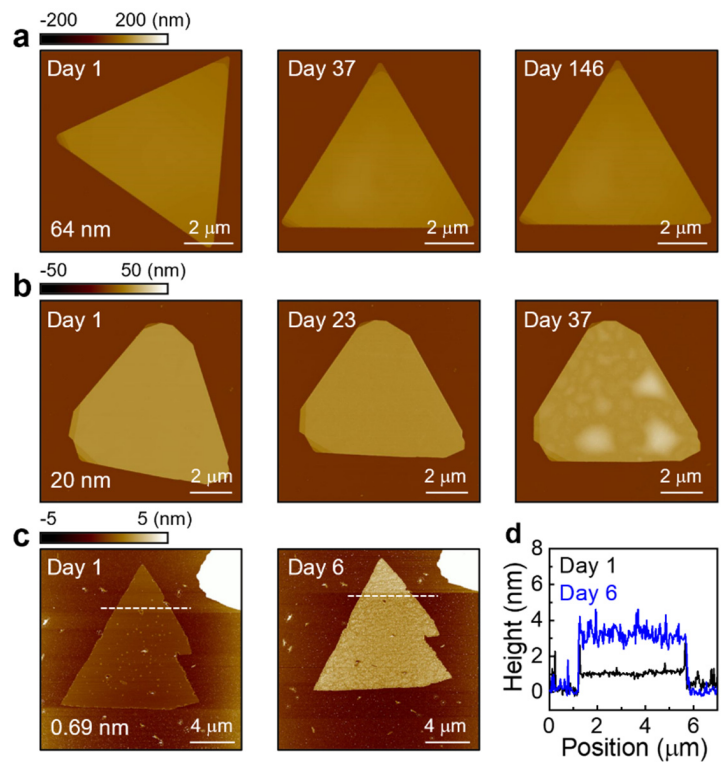

**Figure S2.** Room temperature stability of CrCl$_3$ flakes. a) AFM images of a 64 nm thick CrCl$_3$ flake on mica taken on Day 1, 37, 146 after sample growth. b) AFM images of a 20 nm thick CrCl$_3$ flake on mica taken on Day 1, 23, and 37 after sample growth. c) AFM images of a 1L CrCl$_3$ flake on mica taken on Day 1 and 6 after sample growth, with d) the corresponding height profiles along the dashed lines. This sample is the flake F3 marked in Figure S1a.

## 3. Sample Damage upon TEM and Raman Measurements

The CrCl$_3$ samples can be easily damaged upon exposure to high-energy electron beam and high-power laser excitation. **Figure S3** shows the sample images after TEM and Raman measurements, where the damaged spots are clearly visible. To ensure the collected data quality being preserved during measurements, we have reduced the exposure time and used minimal laser power, which result in very weak signals from thin flakes. For example, as shown in Figure 3a, the Raman response for flakes thinner than 20 nm cannot be barely resolved at the laser power of 0.2 mW. Therefore, our TEM and Raman characterizations are focused on relatively thick flakes.

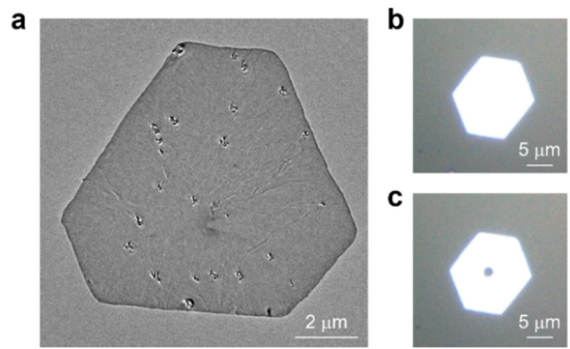

**Figure S3**. Sample damage upon TEM and Raman measurements. a) TEM image of a CrCl$_3$ flake showing damaged spots after electron beam exposure. b-c) Optical images of a CrCl$_3$ flake before (b) and after (c) Raman measurements.

## 4. Element Analysis

**Figure S4** shows the energy dispersive x-ray spectroscopy (EDS) characterization of the flake discussed in Figures 2d-f in the main text. The highest peaks are associated with Si and Au, which come from the Au coated Si substrate. In the expanded spectrum, we show the major Cr and Cl peaks, from which we analyze the atomic percentage of the elements.

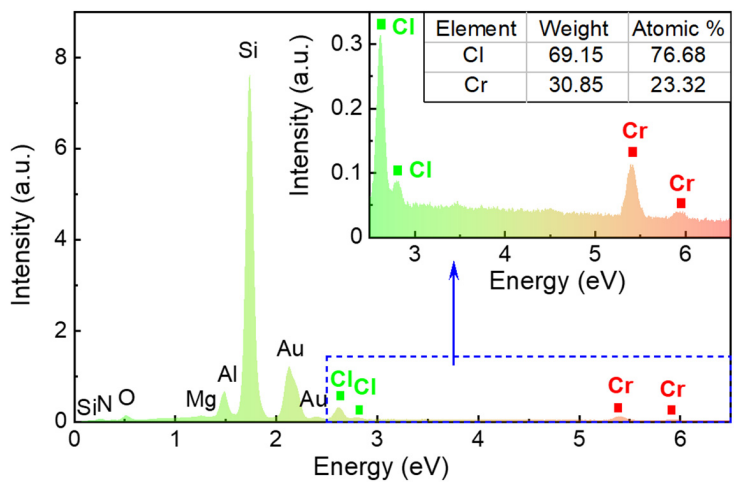

**Figure S4.** EDS spectrum of a 150 nm thick CrCl$_3$ flake. Inset: Expanded view showing the Cl and Cr peaks.

## 5. Polarized Raman Analysis of Crystalline Orientation

As discussed in Figure 3c, the Raman intensity of the A$_g$ modes reaches maximum in parallel polarization when $\theta = 0°$, namely, $\boldsymbol{g}_s \parallel \boldsymbol{a}$ (**Figure S5**a). Based on the Raman polar mapping, we can identify the crystalline orientation of the sample. Figure S5b shows the assigned crystal axes and facets of the CrCl$_3$ flake discussed in Figures 3b-c. It confirms that the {020} and {110} facets are favored during nucleation and growth, which agrees with our TEM analysis (Figures 2b-c).

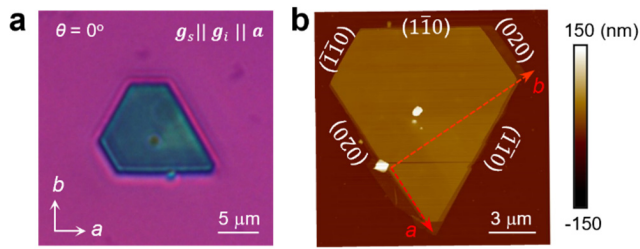

**Figure S5.** Crystalline orientation of CrCl$_3$ flake. a) Optical and b) AFM images of the 43 nm flake discussed in Figures 3b-c, with the crystalline orientation labelled.

## 6. Assembly of h-BN Encapsulated Graphite/CrCl$_3$/Graphite Tunnel Junction

**Figure S6** shows the optical images of the step-by-step process flow for assembling the h-BN encapsulated graphite/few-layer CrCl$_3$/graphite tunnel junction.

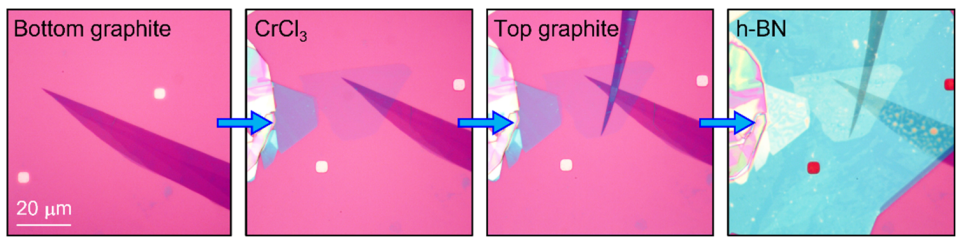

**Figure S6.** Optical images showing the process flow for assembling the h-BN encapsulated graphite/CrCl$_3$/graphite tunnel junction.

## 7. Thickness of CrCl$_3$ Tunnel Barrier

**Figure S7**a shows the optical image of the assembled few-layer CrCl$_3$ tunnel junction device discussed in Figures 4-5 before encapsulation with h-BN. The thickness of the CrCl$_3$ flake has been estimated using calibrated color contrast for CrCl$_3$ on SiO$_2$. As shown in Figure S7b, the color contrast of the tunnel barrier suggests that it is thicker than flake F1 (4-layer) and thinner than flake F2 (8-layer), whose thicknesses are calibrated by AFM (Figure S7b-e). We thus conclude that the thickness of the tunnel barrier is 6±1 layer.

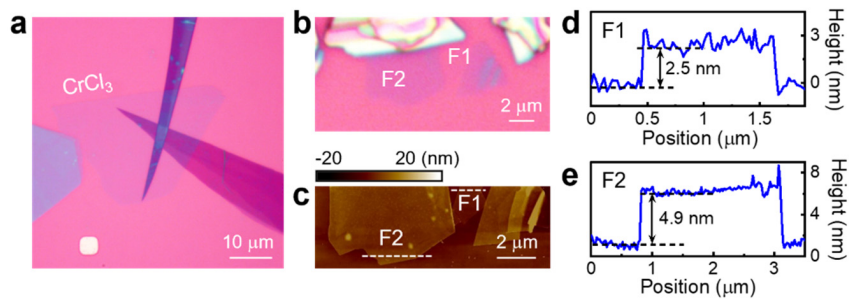

**Figure S7.** Thickness calibration for the CrCl$_3$ tunnel barrier. a) Optical image of the graphite/few-layer CrCl$_3$/graphite device before h-BN encapsulation. b-e) Characterization of few-layer CrCl$_3$ flakes on bare SiO$_2$/Si subsrate for thickness calibration. b) Optical image. c) AFM image, with the height profiles along the dahsed lines for flakes F1 (d) and F2 (e).

## 8. Ambient Stability of h-BN Encapsulated Few-Layer CrCl$_3$ Tunnel Junction

**Figure S8** shows the tunneling *I-V* of the 6-layer CrCl$_3$ tunnel junction device at 300 K taken at various time. At room temperature, there is no prominent change in the *I-V* characteristic, showing excellent ambient stability for over 2 months for the h-BN encapsulated device. This result is discussed in Section 2.3 in the main text.

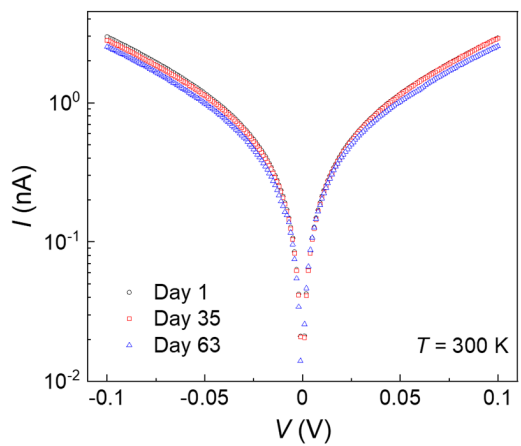

**Figure S8.** Tunneling *I-V* of the 6-layer CrCl$_3$ tunnel junction device at 300 K taken on Day 1, 35, and 63 after fabrication.

25